\documentclass[twocolumn,letter]{jpsj3}
\usepackage{txfonts}
\usepackage{bm}
\usepackage{braket}
\usepackage{color}

\title{Mass-Imbalanced Superconductivity in Effective Two-Channel Kondo Lattice}

\author{Hiroaki Kusunose\thanks{E-mail address: hk@meiji.ac.jp}
}

\inst{Department of Physics, Meiji University, Kawasaki 214-8571, Japan
}

\abst{
We propose that mass-imbalanced superconductivity is realized in an effective two-channel Kondo lattice, and its characteristic property appears in electromagnetic responses such as the Meissner effect.
Starting from an effective two-channel Kondo lattice model as a low-energy effective theory, and approximating it with two mean-field order parameter components in a self-consistent fashion, it is shown that the balance of the two components is sensitively reflected in the magnitude of the Meissner kernel, while thermodynamic properties are little affected by the balance.
This remarkable behavior is understood by the localized character of one partner in the Cooper pair, namely, the effect of the mass imbalance.
We briefly mention the relevance to the huge enhancement of the upper critical field under pressure observed in Pr 1-2-20 systems.
}

\begin{document}
\maketitle

The pairing between fermions is at the heart of superconductivity.
In ordinary metals, a pair is formed by the same kind of conduction electrons on the same Fermi surface with opposite momenta.
In such a situation, the effective mass of each electron in the pair becomes the same order of magnitude, namely, a mass-balanced pair is expected.
Meanwhile, in ultracold fermionic quantum gases, two different atomic species take part in the formation of a Cooper pair, and the mass difference in the pair leads to a variety of interesting phenomena, which have been extensively studied in recent years.~\cite{Sarma63,Barrois77,Liu03,Regal04,Zwielein05,Iskin06,Kasprzak06,Lin06,Wu06,Dao07,Sheehy07,Conduit08,Orso08,Taglieber08,Wille08,Shin08,Guo09,Voigt09,Baarsma10,Diener10,Yoshioka11,Zwerger11,Takemori12,Hanai14,Hanai14a}

In this Letter, we propose that such mass-imbalanced superconductivity can be realized in electronic systems described by an effective two-channel Kondo lattice, and its characteristic property could appear in electromagnetic responses such as the Meissner effect.~\cite{He06,He06a}
On the basis of the effective two-channel Kondo lattice model with two mean-field (MF) variables, we discuss the interplay between the behavior of the Meissner kernel and the ratio of the two components.
When the two components coexist with equal weight, the Meissner kernel becomes largest, while when one of the two components vanishes, the Meissner kernel collapses due to the localized character of one partner in the Cooper pair.
In contrast, thermodynamic properties are little affected by the balance of the two components.

The two-channel Kondo lattice (TCKL) model is one of the standard models in $f$-electron systems, particularly systems having orbital degeneracy.~\cite{Cox98}
For instance, it has been argued extensively that the low-energy phenomena in Pr$T_{2}X_{20}$ ($T$=Ir, Rh, V, Ti; $X$=Zn, Al) can be described by the TCKL model, in which an electric quadrupole plays an important role.~\cite{Onimaru16}
The Hamiltonian of the TCKL model is given by
\begin{equation}
H_{\rm TCKL}=H_{0}+J\sum_{i}(\bm{s}_{i1}+\bm{s}_{i2})\cdot\bm{S}_{i},
\end{equation}
where $H_{0}$ is the kinetic energy of conduction ($c$) electrons, and $\bm{s}_{il}=\sum_{\sigma\sigma'}c_{il\sigma}^{\dagger}(\bm{\sigma}_{\sigma\sigma'}/2)c_{il\sigma'}^{}$ and $\bm{S}_{i}$ denote the $c$-electron spin in channel $l=1,2$ and the localized $f$-electron spin at lattice site $i$.

Analysis using the dynamical mean-field theory (DMFT) has shown that the model exhibits intriguing diagonal and off-diagonal composite orders~\cite{Hoshino11,Hoshino13,Hoshino14,Flint14,Flint14a} such as $\braket{\bm{s}_{i1}\cdot\bm{S}_{i}}-\braket{\bm{s}_{i2}\cdot\bm{S}_{i}}\ne0$ in the electron density ($n_{\rm e}$)-temperature ($T$) phase diagram, in addition to conventional antiferromagnetic and ferromagnetic orders such as $\braket{\bm{S}_{i}}\ne0$.
Among these orders, the composite superconductivity involving the localized $f$-electron spin with staggered center-of-mass momentum $\bm{Q}$ is fascinating, which appears in a wide region of the phase diagram.~\cite{Hoshino14}
This peculiar superconducting state can also be regarded as an ``odd-frequency'' pairing of conduction electrons in both the spin and channel singlet sectors if we integrate out the $f$-electron degrees of freedom.

\begin{figure}[t]
\centering{
\includegraphics[width=8.5cm]{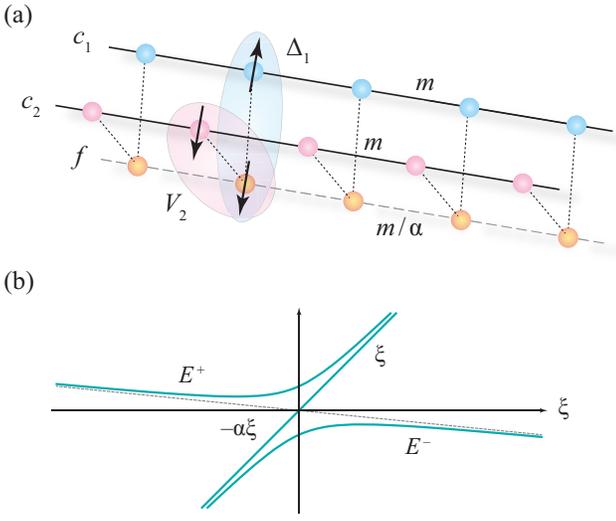}
}
\caption{(Color online) Schematic illustration of the effective two-channel Kondo lattice. (a) Mean-field picture of the composite superconductivity: the Cooper pair of the $f$ and $c$ electrons in the channel 1, and the particle-hole pair (emergent hybridization) of the $f$ and $c$ electrons in channel 2. The effective mass of $c$ ($f$) electrons is given by $m$ ($m/\alpha$). (b) Three energy bands in the composite ordered phase.}
\label{composite}
\end{figure}

Hoshino proposed an effective MF description for the superconducting state~\cite{Hoshino14a,Flint14,Flint14a} with a fictitious pseudofermion $f_{i\sigma}$ representing the $f$-electron spin $\bm{S}_{i}$ in the low-energy effective theory.
The MF description was verified by reproducing the essential features of one-particle properties obtained by DMFT calculation.
On the basis of the MF Hamiltonian with two order parameter components, $W_{1}$ (which we denote as $\Delta_{1}$) and $V_{2}$ (see Fig.~\ref{composite}), he investigated the fundamental properties of the Meissner kernel.
However, an explicit form of a Hamiltonian that leads to the MF description was not given, and the self-consistency and stability of the MF solutions were not elucidated in detail.

In this Letter, we perform the MF analysis in a fully self-consistent fashion.
To this end, let us introduce the following effective Hamiltonian to be treated by the MF approximation:
\begin{equation}
H=\sum_{\bm{k}}\sum_{\sigma}^{\uparrow,\downarrow}\xi_{\bm{k}}\left(\sum_{l}^{1,2}c_{\bm{k}l\sigma}^{\dagger}c_{\bm{k}l\sigma}^{}-\alpha f_{\bm{k}\sigma}^{\dagger}f_{\bm{k}\sigma}^{}\right)+H_{\rm int}.
\end{equation}
The channel and spin indices, $l=1,2$ and $\sigma=\uparrow,\downarrow$, should be regarded appropriately as ($+1,-1$) if necessary in the following equations.
Here, we assume that the kinetic energy satisfies the relation $\xi_{\bm{k}}=-\xi_{\bm{Q}-\bm{k}}$, which ensures the stability of the superconducting state with the staggered ordering vector $\bm{Q}$.
In addition to this property, we assume an isotropic and continuum system for simplicity.
Moreover, we introduce a small finite bandwidth $\alpha\xi_{\bm{k}}$ of the pseudofermion\cite{Coleman15} (called the ``$f$-electron'' hereafter) $f_{\bm{k}\sigma}$ to keep track of the effect of the large but finite effective mass of the $f$ electron, $m/\alpha$, where the effective mass of $c$ electrons is given by $1/m=\partial^{2}\xi_{\bm{k}}/\partial\bm{k}^{2}$ at the Fermi energy.
It seems to be natural to introduce a finite bandwidth considering that the TCKL model was originally derived from the corresponding Anderson lattice model, in which $f$ electrons are often described by weak dispersive bands.

We consider the effective interaction in the form
\begin{multline}
H_{\rm int}=
-\frac{1}{2N_{0}}\sum_{\bm{k}\bm{k}'}\sum_{l\sigma}\biggl[
\frac{U}{1-r}\,f_{\bm{Q}-\bm{k}'\bar{\sigma}}^{\dagger}c_{\bm{k}'l\sigma}^{\dagger}c_{\bm{k}l\sigma}^{}f_{\bm{Q}-\bm{k}\bar{\sigma}}^{}
\\
+\frac{U}{1+r}\,f_{\bm{k}'\sigma}^{\dagger}c_{\bm{k}'l\sigma}^{}c_{\bm{k}l\sigma}^{\dagger}f_{\bm{k}\sigma}^{}
\\
+\eta U\biggl\{(l\sigma)
f_{\bm{k}'\bar{\sigma}}^{\dagger}c_{\bm{k}'\bar{l}\bar{\sigma}}^{}c_{\bm{k}l\sigma}^{}f_{\bm{Q}-\bm{k}\bar{\sigma}}^{}
+{\rm h.c.}
\biggr\}
\biggr],
\label{hint}
\end{multline}
where $N_{0}$ is the number of lattice sites.
The description for each term in $H_{\rm int}$ is given below.
For this interaction, we introduce two types of order parameter,
\begin{subequations}
\begin{align}
&
\Delta_{l}=\frac{U}{1-r}\frac{1}{N_{0}}\sum_{\bm{k}}\sigma\braket{c_{\bm{k}l\sigma}^{}f_{\bm{Q}-\bm{k}\bar{\sigma}}^{}},
\\&
V_{l}=\frac{U}{1+r}\frac{1}{N_{0}}\sum_{\bm{k}}\braket{c_{\bm{k}l\sigma}^{\dagger}f_{\bm{k}\sigma}^{}},
\end{align}
\end{subequations}
where $\Delta_{l}$ is a singlet pair with finite center-of-mass momentum $\bm{Q}$, and $V_{l}$ is the particle-hole pair (emergent hybridization) between the $c$ electron in channel $l$ and the $f$ electron.
A schematic illustration of the composite superconducting state and the dispersion relations in the ordered state is shown in Fig.~\ref{composite}(a).

Now, the meaning of each term in Eq.~(\ref{hint}) is apparent.
The first and second terms favor the singlet particle-particle and particle-hole pairs, respectively, where we have introduced the asymmetric factor $r$ to discriminate the attractive interactions for these two components.
Note that for $r=0$, the effective interaction is invariant under the particle-hole transformation
$f_{\bm{k}\sigma}^{}\to \sigma f_{\bm{Q}-\bm{k}\bar{\sigma}}^{\dagger}$, which exchanges the roles of $\Delta_{l}$ and $V_{l}$.
The third term is the attraction between $\Delta_{l}$ and $V_{\bar{l}}$.
This term should not appear in the normal state since it does not conserve the number of particles.
Nevertheless, a state with coexisting $\Delta_{l}$ and $V_{\bar{l}}$ is indeed realized as shown by the DMFT calculation,~\cite{Hoshino14}
which may be understood by a higher-order coupling term between $\Delta_{l}$ and $V_{\bar{l}}$ in the ordered state.
To reproduce the coexistence at the level of the MF approximation, the direct attraction, such as that given by the third term, mimics such a coupling effect.
It should be emphasized that the interaction in Eq.~(\ref{hint}) is meaningful only when the MF approximation is adopted.
It is sufficient for our purpose to elucidate the relative deviation from the most coexisting state given at $r=0$ as follows.

In terms of these order parameters, the free energy per site measured from that of the normal state is given by
\begin{equation}
F=
\frac{g(\theta,\phi)\Delta^{2}}{U}
-\frac{4T}{N_{0}}\sum_{\bm{k}}
\ln\left[
\frac{(1+e^{-E_{\bm{k}}^{+}/T})(1+e^{-E_{\bm{k}}^{-}/T})}{(1+e^{-\xi_{\bm{k}}/T})(1+e^{\alpha\xi_{\bm{k}}/T})}
\right],
\end{equation}
where $E_{\bm{k}}^{\pm}=[(1-\alpha)\xi_{\bm{k}}\pm \sqrt{(1+\alpha)^{2}\xi_{\bm{k}}^{2}+\Delta^{2}}]/2$ [see Fig.~\ref{composite}(b)].
Note that there is another branch, $\xi_{\bm{k}}$, which remains unchanged through the phase transition, and hence it does not contribute to the change in the free energy, although it gives a gapless feature in the thermodynamics in the superconducting state.
Here, we have factorized the order parameters\cite{factorize} in terms of $\Delta$, $\phi$, and $\theta$ as
\begin{subequations}
\begin{align}
&
(\Delta_{1},V_{2})=\frac{\sqrt{1+\eta^{2}}}{1-\eta^{2}}\Delta \cos\phi\,[\sin(\theta-\theta_{0}),\cos(\theta+\theta_{0})],
\\&
(\Delta_{2},V_{1})=\frac{\sqrt{1+\eta^{2}}}{1-\eta^{2}}\Delta \sin\phi\,[\sin(\theta+\theta_{0}),\cos(\theta-\theta_{0})],
\end{align}
\label{ops}
\end{subequations}
where $\tan\theta_{0}=\eta$ [it is equivalent to $\theta_{0}\equiv \theta^{*}(-r_{0},\eta)$, see below].
The phase of $\Delta$ is chosen as real without loss of generality.
The $\phi$ and $\theta$ dependences appear only through the effective interaction strength introduced as
\begin{multline}
g(\theta,\phi)=\frac{1}{1-\eta^{2}}\biggl(
1+\frac{2\eta^{2}r^{2}}{1-\eta^{2}}+r\cos(2\theta)
\\
-\eta\left(1+\frac{r^{2}}{r_{0}}\right)\sin(2\theta)\cos(2\phi)
\biggr),
\end{multline}
where $r_{0}=(1-\eta^{2})/(1+\eta^{2})$.

For $\eta=r=0$, the free energy is independent of $\phi$ and $\theta$, and the order parameters minimizing $F$ are degenerate for arbitrary values of $\phi$ and $\theta$.
This limit corresponds to the subgroup of the SO(5) symmetry group\cite{so5} as argued in the literature.~\cite{Hoshino11,Hoshino13,Hoshino14,Hoshino14a}
For $\eta>0$, the largest $T_{\rm c}$ is obtained for $\phi=0$ and $\theta=\theta^{*}$, which is given explicitly by
\begin{equation}
\theta^{*}(r,\eta)=\frac{1}{2}\arctan\left[\frac{r}{\eta(1+r^{2}/r_{0})}\right]+\frac{\pi}{4}.
\end{equation}
Note that it satisfies $\theta^{*}(r,\eta)+\theta^{*}(-r,\eta)=\pi/2$.
As the case of $\phi=\pi/2$ is essentially equivalent to that of $\phi=0$, we restrict ourselves to the case of $\phi=0$.
In what follows, we consider $\eta>0$, $\phi=0$ ($\Delta_{2}=V_{1}=0$), and we denote $g^{*}\equiv g(\theta^{*},0)$.
Note that the superconducting order parameter $\Delta_{1}$ vanishes at $r=-r_{0}$ ($\theta^{*}=\theta_{0}$), while $V_{2}=0$ for $r=+r_{0}$ ($\theta^{*}=\pi/2-\theta_{0}$).
For $r=0$, we have $\theta^{*}=\pi/4$ with $\Delta_{1}=V_{2}$.

The self-consistent equation is obtained by differentiating $F(\phi=0,\theta=\theta^{*})$ with respect to $\Delta$ to obtain
\begin{equation}
\frac{g^{*}}{U}=\frac{1}{N_{0}}\sum_{\bm{k}}\frac{f(E_{\bm{k}}^{-})-f(E_{\bm{k}}^{+})}{E_{\bm{k}}^{+}-E_{\bm{k}}^{-}},
\end{equation}
where $f(x)=1/(e^{x/T}+1)$ is the Fermi--Dirac distribution function.
This equation is similar to that for a density-wave order rather than the BCS gap equation far below $T_{\rm c}$.
However, for $\alpha=0$, the linearized equation is reduced to the same form as the BCS gap equation with the attractive interaction $U/g^{*}$.
At $T=0$, the self-consistent equation can be solved analytically to obtain
\begin{equation}
\Delta(0)=\frac{(1+\alpha)D}{\sinh[(1+\alpha)g^{*}/2\rho_{\rm F}U]},
\end{equation}
where $D$ is half of the $c$-electron bandwidth and $\rho_{\rm F}$ is the $c$-electron density of states per spin and channel at the Fermi energy.
We use the cutoff $D$ as the unit of energy and fix $\rho_{\rm F}U=0.6$ and $\eta=0.1$ throughout this paper.

\begin{figure}[t]
\centering{
\includegraphics[width=8.5cm]{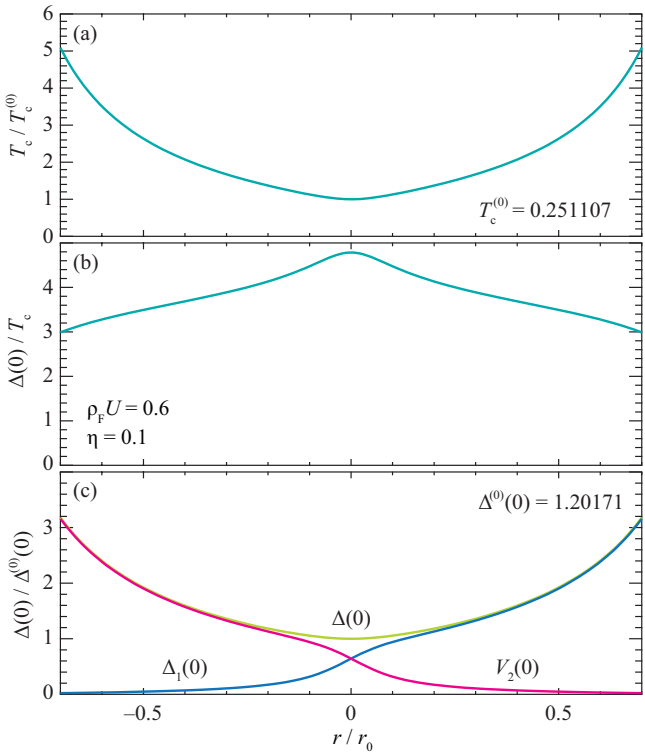}
}
\caption{(Color online) Asymmetric factor $r$ dependences for $\rho_{\rm F}U=0.6$, $\eta=0.1$, and $\alpha=0$: (a) transition temperature $T_{\rm c}$, (b) $\Delta(0)/T_{\rm c}$ at $T=0$, and (c) two order parameter components at $T=0$, $\Delta_{1}(0)$ and $V_{2}(0)$, together with the magnitude $\Delta(0)$. Note that at $r=\pm r_{0}$, one of the components, $V_{2}$ or $\Delta_{1}$, vanishes. The dependences for $\alpha=10^{-2}$ (not shown) are almost unchanged. The values with the superscript $(0)$ indicate the values for $\alpha=r=0$.}
\label{rdep}
\end{figure}

First, we discuss the $r$ dependences.
Figure~\ref{rdep} shows the $r$ dependences of $T_{\rm c}$, $\Delta(0)/T_{\rm c}$, $\Delta(0)$, $\Delta_{1}(0)$, and $V_{2}(0)$ at $T=0$ for $\alpha=0$.
The values for $\alpha=r=0$ are indicated by the superscript $(0)$.
The behaviors for $\alpha=10^{-2}$ (not shown) are almost unchanged.
As shown in Fig.~\ref{rdep}(c), one of the gap components is strongly suppressed as $r$ deviates from $r=0$, but the magnitude of the gap $\Delta(0)$ is always finite, which determines the thermodynamic properties since the quasiparticle bands depend only on $\Delta$.

Next, we elucidate the nature of the Meissner kernel.
To obtain an expression for the Meissner kernel in the London limit ($\bm{q}\to0$), we utilize the Nambu representation,
\begin{align}
C_{\bm{k}}^{\dagger}
&=
\begin{pmatrix}
c^{\dagger}_{\bm{k}1\sigma}
\,\,
c^{}_{\bm{Q}-\bm{k}2\bar{\sigma}}
\,\,
f^{}_{\bm{Q}-\bm{k}\bar{\sigma}}
\,\,\bigl|\,\,
c^{}_{\bm{Q}-\bm{k}2\bar{\sigma}}
\,\,
c_{\bm{k}2\sigma}^{\dagger}
\,\,
f^{\dagger}_{\bm{k}\sigma}
\end{pmatrix}
\cr&
=
\begin{pmatrix}
C_{A\bm{k}\sigma}^{\dagger}\,\, C_{B\bm{k}\sigma}^{\dagger}
\end{pmatrix}.
\end{align}
The Nambu space can be decoupled into two parts, $A$ and $B$, for $\phi=0$, and the two parts give essentially the same contributions to the physical quantities.
Thus, we consider only part $A$.
The matrix of the Green's function, $G(\xi,i\omega_{n})=-\int_{0}^{1/T}d\tau\,e^{i\omega_{n}\tau}\braket{T_{\tau}\,C^{}_{A\bm{k}\sigma}(\tau)C^{\dagger}_{A\bm{k}\sigma}}$, where $\omega_{n}=\pi T(2n+1)$ is the fermionic Matsubara frequency, is given by
\begin{equation}
G(\xi,z)=\frac{1}{W(\xi,z)}
\begin{pmatrix}
y-d_{\rm c}^{2} & d_{\rm c}d_{\rm s} & (z-\xi)d_{\rm s} \\
d_{\rm c}d_{\rm s} & y-d_{\rm s}^{2} & (z-\xi)d_{\rm c} \\
(z-\xi)d_{\rm s} & (z-\xi)d_{\rm c} & (z-\xi)^{2}
\end{pmatrix}.
\end{equation}
Here, $y=(z-\xi)(z+\alpha\xi)$, $d_{\rm c}=(\Delta/2)\cos\theta^{*}$, $d_{\rm s}=(\Delta/2)\sin\theta^{*}$, and the determinant is $W=(z-\xi)(y-\Delta^{2}/4)$.
Note that $G(\xi,z)$ directly depends on $(r,\eta)$ only through $\theta^{*}(r,\eta)$.

By performing the standard procedure,~\cite{AGD63,Altland10} we express the Meissner kernel in terms of the Green's function as
\begin{equation}
K_{\rm s}(T)=\int_{-D}^{D}d\xi\,T\sum_{n}\biggl( I(\xi,i\omega_{n})-I_{0}(\xi,i\omega_{n}) \biggr),
\end{equation}
where $I$ is given by the component of the Green's function as
\begin{equation}
I(\xi,z)=G_{11}^{2}+G_{22}^{2}-2G_{12}^{2}+\alpha^{2}G_{33}^{2}-2\alpha G_{23}^{2}+2\alpha G_{13}^{2},
\end{equation}
and $I_{0}$ is the value for $\Delta=0$.
We have used the relation $1/\rho_{\rm F}=2m\varv_{\rm F}^{2}/3$.
The superconducting current is related to the vector potential as $\bm{j}=-(e^{2}/mc)K_{\rm s}(T)\bm{A}$.

By using the fact that $K_{\rm s}(T)$ must vanish at $r=-r_{0}$ ($\theta^{*}=\theta_{0}$), where $\Delta_{1}=0$, we can eliminate the contribution from $I_{0}$ by using the value $I$ at $\theta_{0}$, and we finally obtain the expression
\begin{multline}
K_{\rm s}(T)=
K_{\rm s}^{a}(T)\biggl\{
\sin^{2}(2\theta^{*})-\sin^{2}(2\theta_{0})
\biggr\}
\\
+K_{\rm s}^{b}(T)\biggl\{
\sin^{2}\theta^{*}-\sin^{2}\theta_{0}
\biggr\},
\label{mk}
\end{multline}
where
\begin{subequations}
\begin{align}
&
K_{\rm s}^{a}(T)=-\frac{\Delta^{4}}{8}\int_{0}^{D}d\xi\,T\sum_{n}\frac{1}{W^{2}(\xi,i\omega_{n})},
\\&
K_{\rm s}^{b}(T)=2\alpha\Delta^{2}\int_{0}^{D}d\xi\,T\sum_{n}\frac{(i\omega_{n}-\xi)^{2}}{W^{2}(\xi,i\omega_{n})}.
\end{align}
\label{mkp}
\end{subequations}
Note that $K_{\rm s}^{b}(T)$ has a prefactor $\alpha$, and hence it vanishes for $\alpha=0$.
The Matsubara summation in Eq.~(\ref{mkp}) can be carried out by using a contour integral, but it is not shown here because the explicit expressions are somewhat complicated.

\begin{figure}[t]
\centering{
\includegraphics[width=8.5cm]{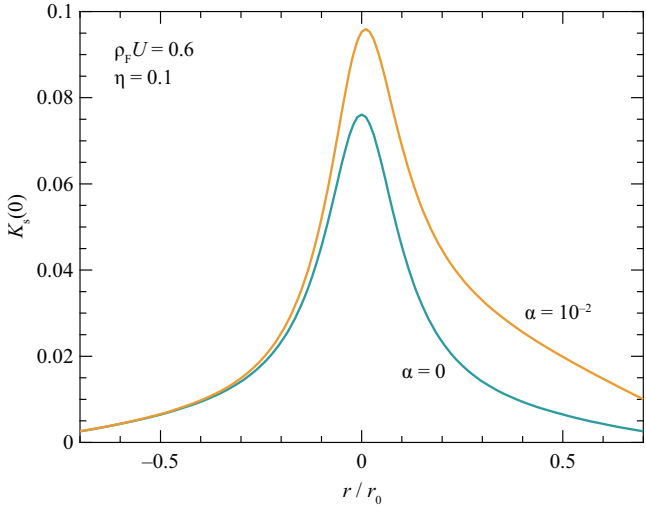}
}
\caption{(Color online) Asymmetric factor $r$ dependence of $K_{\rm s}(0)$ at $T=0$ for $\rho_{\rm F}U=0.6$, $\eta=0.1$. Note that $K_{\rm s}$ is symmetric for $\alpha=0$.}
\label{ks}
\end{figure}

Equation~(\ref{mk}) indicates that $K_{\rm s}(T)$ vanishes at $r=-r_{0}$ ($\theta^{*}=\theta_{0}$) owing to the sine factors, as it should.
With increasing $r$ from $r=-r_{0}$, $K_{\rm s}(T)$ increases and it reaches a maximum around $r=0$ ($\theta^{*}=\pi/4$), where two order parameters coexist most constructively as shown in Fig.~\ref{rdep}(c).
With a further increase in $r$, $K_{\rm s}(T)$ decreases again and it even vanishes at $r=+r_{0}$ ($\theta^{*}=\pi/2-\theta_{0}$) for $\alpha=0$, where $V_{2}=0$ and only the pairing between $c_{\bm{k}1\sigma}$ and $f_{\bm{Q}-\bm{k}\bar{\sigma}}$ is realized.
In this case, it cannot carry the supercurrent since one of the partner of the pair is completely localized with zero velocity.
For small but finite $\alpha$, the $f$ electron acquires a small velocity, which makes $K_{\rm s}(T)$ small but finite.
Thus, the smallness of $K_{\rm s}(T)$ at large $|r|$ reflects the localized character of the $f$ electron, namely, the mass imbalance in the pairing.
The typical behavior of $K_{\rm s}(0)$ at $T=0$ as a function of $r$ is shown in Fig.~\ref{ks}.

Performing the $\xi$ integration in Eq.~(\ref{mkp}) at $T=0$, we obtain the explicit expressions
\begin{subequations}
\begin{align}
&
K_{\rm s}^{a}(0)=1-\frac{x_0(3+4x_0^{2}-4x_0\sqrt{1+x_0^{2}})}{(1+\alpha)\sqrt{1+x_0^{2}}},
\\&
K_{\rm s}^{b}(0)=\frac{4\alpha}{1+\alpha}\frac{x_0}{\sqrt{1+x_0^{2}}},
\end{align}
\end{subequations}
where $x_0=(1+\alpha)D/\Delta(0)$.
In the large-cutoff limit, we have the asymptotic forms
\begin{subequations}
\begin{align}
&
K_{\rm s}^{a}(0)\sim \frac{\alpha}{1+\alpha}+\frac{\Delta^{4}(0)}{8(1+\alpha)^{5}D^{4}},
\\&
K_{\rm s}^{b}(0)\sim \alpha\left[\frac{4}{1+\alpha}-\frac{2\Delta^{2}(0)}{(1+\alpha)^{3}D^{2}}\right],
\end{align}
\end{subequations}
which strongly depend on $\Delta(0)/D$, and they vanish at $D\to\infty$ for $\alpha=0$.
Therefore, the Meissner kernel becomes very small for $\Delta(0)/D\ll1$.

\begin{figure}[t]
\centering{
\includegraphics[width=8.5cm]{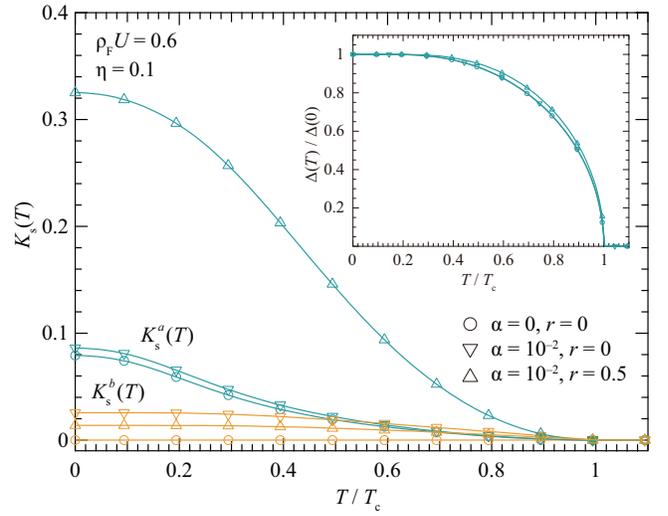}
}
\caption{(Color online) $T$ dependences of the two distinct contributions to the Meissner kernel. The inset shows the $T$ dependence of the gap magnitude $\Delta(T)$ normalized by $\Delta(0)$.}
\label{tdep}
\end{figure}

Finally, we discuss the $T$ dependences.
Figure \ref{tdep} shows the $T$ dependences of the two distinct contributions to the Meissner kernel, $K_{\rm s}^{a}(T)$ and $K_{\rm s}^{b}(T)$, and the gap magnitude $\Delta(T)$ in the inset.
$K_{\rm s}^{a}(T)$ and $K_{\rm s}^{b}(T)$ show monotonically increase with decreasing $T$, similarly to the ordinary Meissner kernel.
Thus, the possible marked change in the Meissner kernel originates from the $r$ dependence, i.e., the change in the ratio of the attractive interactions.
The normalized gap magnitude as a function of $T/T_{\rm c}$ shows almost the same $T$ dependence, as shown in the inset.
The absolute magnitude of $\Delta$ changes according to Fig.~\ref{rdep} and has moderately weak $r$ dependence.

We briefly mention the experimental relevance.
The Pr-based cubic system Pr$T_{2}X_{20}$ has attracted much attention because of its peculiar behaviors inherent from its orbital degeneracy.~\cite{Onimaru16}
It has been argued that the low-energy properties of Pr 1-2-20 systems can be described by the TCKL model.
In particular, PrTi$_{2}$Al$_{20}$ successively undergoes quadrupole ordering and exhibits superconductivity at $T_{\rm Q}=2$ K and $T_{\rm c}=0.2$ K, respectively, at ambient pressure.~\cite{Matsubayashi12,Matsubayashi14}
Upon applying pressure, $T_{\rm c}$ is enhanced to as high as $T_{\rm c}=1.1$ K at $P=8.7$ GPa, namely, $T_{\rm c}$ becomes about five times higher under pressure.
In contrast, the upper critical field extrapolated to $T=0$ is enhanced dramatically from $B_{\rm c2}(0)=6.3$ mT at ambient pressure to $3.5$ T at $P=8.7$ GPa (by about 560 times).
As in ordinary metals, the upper critical field is roughly proportional to $T_{\rm c}^{2}$, and the observed enhancement is one order of magnitude larger.
This puzzle may be resolved by considering the role of the mass imbalance discussed in this paper.
In other words, the coherence length is related as $\xi_{0}\propto K_{\rm s}(0)$, and then the upper critical field is scaled by $B_{\rm c2}(0)\propto \xi_{0}^{-2}\propto [K_{\rm s}(0)]^{-2}$.
For example, by comparing the values at $r=0$ and $r=\pm 0.7r_{0}$ in Figs.~\ref{rdep} and \ref{ks}, the enhancement of $T_{\rm c}$ is about five times, while that of $[K_{\rm s}(0)]^{-2}$ becomes as large as $10^{2}$-$10^{3}$ due to the effect of the mass imbalance.
Although an explicit estimate of the change in the attractive interaction is not available at moment, further experimental and theoretical investigations will shed light on the interplay between the marked enhancement of $B_{\rm c2}$ and the character of the mass imbalance.

In summary, we propose that mass-imbalanced superconductivity can be realized in the effective two-channel Kondo lattice, where the interplay between two order parameters ($\Delta_{1}$ and $V_{2}$) gives rise to a marked change in the Meissner kernel with a moderate change in the thermodynamic properties.
The puzzle of the huge enhancement of the upper critical field observed in PrTi$_{2}$Al$_{20}$ under pressure may be resolved by considering the effect of the mass imbalance.
Further experimental and theoretical investigations are highly desirable to reveal this peculiar mass-imbalanced superconductivity, which is not expected in ordinary metals.

The author would like to thank S. Hoshino, M. Matsumoto, M. Koga, K. Miyake, Y. Yanagi, Y. Ohashi, K. Izawa, T. Onimaru, Y. Kato, and K. Matsubayashi for fruitful discussions.
This work was supported by JSPS KAKENHI Grant Numbers 15K05176 and 15H05885 (J-Physics).


\begin{thebibliography}{99}
\bibitem{factorize} We first introduce $\Delta_{l}'=\Delta_{l}+l\eta V_{-l}$ and $V_{l}'=V_{l}-l\eta\Delta_{-l}$ and they are parameterized as $(\Delta_{1}',V_{2}')=\Delta\cos\phi\,(\sin\theta,\cos\theta)$ and $(\Delta_{2}',V_{1}')=\Delta\sin\phi\,(\sin\theta,\cos\theta)$. Inverting these relations with respect to $V_{l}$ and $\Delta_{l}$, we obtain Eq.~(\ref{ops}).
\bibitem{so5} In addition to $V_{l}$ and $\Delta_{l}$ at $\alpha=\eta=0$, the transverse components of the composite order parameters constitute generators of the SO(5) symmetry group\cite{Hoshino14a}, which have been ignored here at the level of the MF treatment.
\end{thebibliography}
\end{document}